\documentclass[12pt]{article}
\usepackage{graphicx}
\usepackage{amsmath}
\def \sd {\mathbf{D}}

\def \be {\mathbf{E}}

\def \br {\mathbf{r}}

\def \bx {\mathbf{x}}

\def \bp {\mathbf{p}}

\newcommand{\Tint}[1]{{\hbox{$\sum$}\!\!\!\!\!\!\!\int\,}_{\!\!\!\!\raise-0.9ex\hbox{$\scriptstyle{#1}$}}}

\def\siml{{\ \lower-1.2pt\vbox{\hbox{\rlap{$<$}\lower6pt\vbox{\hbox{$\sim$}}}}\ }}
\def\simg{{\ \lower-1.2pt\vbox{\hbox{\rlap{$>$}\lower6pt\vbox{\hbox{$\sim$}}}}\ }}
\def \sd {\mathbf{D}}

\def \be {\mathbf{E}}

\def \br {\mathbf{r}}

\def \bx {\mathbf{x}}

\def \bp {\mathbf{p}}

\def \als {\alpha_{\mathrm{s}}}

\def \m2   {\mu^{2 \epsilon}}

\def\siml{{\ \lower-1.2pt\vbox{\hbox{\rlap{$<$}\lower6pt\vbox{\hbox{$\sim$}}}}\ }}
\def\simg{{\ \lower-1.2pt\vbox{\hbox{\rlap{$>$}\lower6pt\vbox{\hbox{$\sim$}}}}\ }}
\def\lqcd{\Lambda_\mathrm{QCD}}

\def\pbnr{}
\def\speaker{Jacopo Ghiglieri}
\def\onbehalfof{}
\def\title{A brief review of the theory of charmonium suppression in heavy ion collisions}
\def\affiliation{Physics Department\\
McGill University, Montr\'eal, Canada}
\def\support{The speaker is supported by the Natural Sciences and Engineering Research Council of Canada and by an Institute of Particle Physics Theory Fellowship. }

\newcommand\pubnumber{\pbnr}
\newcommand\pubdate{\today}
\def\Title#1{\begin{center} {\Large #1 } \end{center}}
\def\Author#1{\begin{center}{ \sc #1} \end{center}}

\newcommand{\OnBehalf}[1]{\sbox0{#1}\ifdim\wd0=0pt
        {}
	\else
	{\\on behalf of #1}
	\fi}
\newcommand{\SupportedBy}[1]{\sbox0{#1}\ifdim\wd0=0pt
        {}
	\else
	{\footnote{#1}}
	\fi}
\def\Address#1{\begin{center}{ \it #1} \end{center}}

\newcommand\pubblock{\includegraphics[width=5cm]{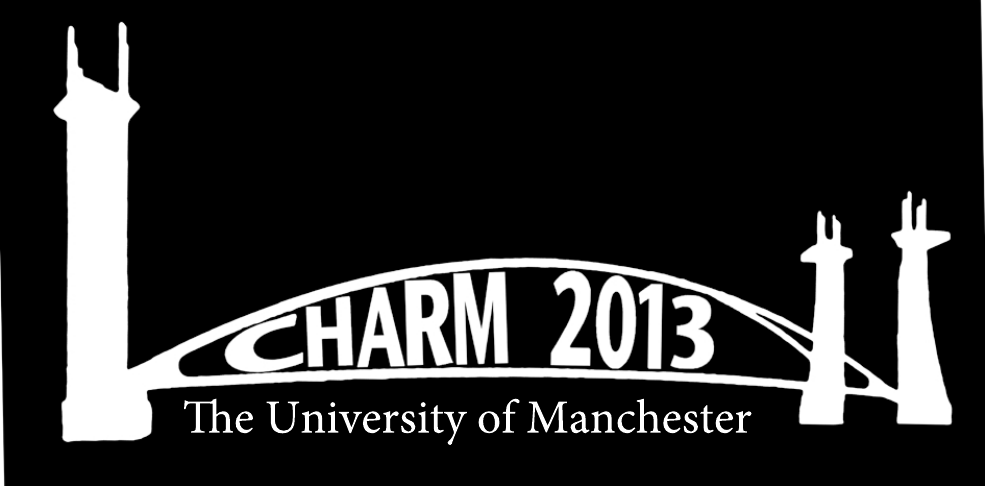}\hfill{\begin{tabular}{l} \pubnumber\\
         \pubdate  \end{tabular}}}
\newenvironment{Abstract}{\begin{quotation}  }{\end{quotation}}
\newenvironment{Presented}{\begin{quotation} \begin{center} 
             PRESENTED AT\end{center}\bigskip 
      \begin{center}\begin{large}}{\end{large}\end{center} \end{quotation}}

\def\venue{The 6$^{th}$ International Workshop on Charm Physics\\
(CHARM 2013)\\
Manchester, UK,  31 August -- 4 September, 2013}

\def\beq{\begin{equation}}
\def\eeq#1{\label{#1}\end{equation}}
\def\eeqn{\end{equation}}

\def\beqa{\begin{eqnarray}}
\def\eeqa#1{\label{#1}\end{eqnarray}}
\def\eeqan{\end{eqnarray}}

\let\bar=\overbar

\def\Dslash{\not{\hbox{\kern-4pt $D$}}}
\def\dslash{\not{\hbox{\kern-2pt $\del$}}}

\def\msb{{\bar{\ssstyle M \kern -1pt S}}}

\begin{document}
\begin{titlepage}
\pubblock

\vfill
\Title{\title}
\vfill
\Author{\speaker\SupportedBy{\support}\OnBehalf{\onbehalfof}}
\Address{\affiliation}
\vfill
\begin{Abstract}
A brief overview of the theory of charmonium suppression in heavy ion collisions
is presented. In particular I will concentrate on the effects caused by the
hot, deconfined medium and on the effort to treat them using field-theoretical, QCD-based
techniques, such as lattice QCD and Effective Field Theories.
\end{Abstract}
\vfill
\begin{Presented}
\venue
\end{Presented}
\vfill
\end{titlepage}
\def\thefootnote{\fnsymbol{footnote}}
\setcounter{footnote}{0}
%


\section{Introduction}
The phase diagram of QCD in the region of vanishing chemical potential is actively 
explored in heavy ion collision experiments at RHIC and LHC. Lattice calculations 
(see \cite{Petreczky:2012rq} for a recent review) predict a crossover transition 
to a larger number of degrees of freedom, typical of a deconfined medium, the Quark-Gluon
Plasma (QGP) for 
temperatures $T$ ranging from 150 to 200 MeV.  On the experimental side, the
characterization of the properties of the medium relies either on its bulk properties,
which find an effective description in hydrodynamics, or  on 
\emph{hard probes}, i.e. energetic particles not in equilibrium
with the medium. 

Heavy quarkonium has been one of the most actively investigated hard probes for the
past 27 years. In 1986 Matsui and Satz \cite{Matsui:1986dk} hypothesized that colour
screening in a deconfined medium would have dissociated the $J/\psi$, resulting
in a suppressed yield in the easily accessible dilepton decay channel and yielding
a striking QGP signature. Experimentally, suppression (or lack thereof) is investigated through the
\emph{nuclear modification factor} $R_{AA}$, defined as the (quarkonium) yield in the 
nucleus-nucleus collision divided by the corresponding one in $pp$, scaled 
by the number of binary collisions.  Heavy ion collision experiments at SPS, RHIC and LHC 
(see \cite{Brambilla:2010cs,Blume_charm} for a review) have indeed reported $R_{AA}<1$.
 Furthermore the LHC results have opened up the frontier of the cleaner bottomonium 
probe with the availability of quality data on the $\Upsilon$ resonances (see 
\cite{CMSupsi} for the latest CMS results). 

The current theoretical understanding is that all stages of a heavy ion collision
contribute to some level to the deviation from simple binary scaling. In the early 
stages one has to address \emph{cold nuclear matter} effects, i.e. those caused by a
confined, nuclear environment. The current understanding is that these effects alone 
cannot explain the observed $R_{AA}<1$. Later, after a fast thermalization (it is believed
to happen in less than 1 fm/c), the system reaches 
an approximate local thermal equilibrium in the deconfined phase.
There one usually speaks of \emph{hot nuclear matter} or \emph{hot medium} effects, such as 
the aforementioned colour screening. The hot medium is rapidly expanding
and cooling down, in a process that is nowadays well described by hydrodynamics 
(see \cite{Huovinen:2013wma} for a recent review), eventually hadronizing into final state particles, in 
a time that is estimated to be at most $\sim10$ fm/c. At this 
stage unbound $c\overline{c}$ pairs might hadronize into a charmonium resonance, in what is known
as \emph{statistical recombination} \cite{statistical}. Any surviving vector state will then decay on
 a timescale that is several order of magnitude longer than those encountered so far.
  It is finally worth remarking that,
when studying the vector ground states $J/\Psi$ and $\Upsilon(1S)$, one needs to take into account
that in a $pp$ collision roughly one half of the observed yield comes from feed-down from excited 
states  \cite{feeddown}. Medium modifications thereto need then to be consistenly taken into account.

In this proceeding we will concentrate on a brief description of some of the issues in the theoretical
description of cold and hot nuclear matter effects. Due to the limited space, we will mostly refer
 to reviews, such as 
\cite{early_reviews,Brambilla:2010cs,Mocsy:2013syh}. The interested reader is invited to
follow up on the original references therein. Finally, recent phenomenological transport models which implement,
in different manners, a number of the aforementioned effects for charmonia and bottomonia can be found in \cite{strickland,Zhao:2010nk,Emerick:2011xu,Song:2011ev,Nendzig:2012cu}.

\section{Cold nuclear matter effects}
A theoretical description of quarkonium production, even in the cleaner $pp$, $pe$ and $e^+e^-$ 
initial states, is a challenging task (see \cite{Brambilla:2010cs,Butenschoen}), which is clearly
made no easier by the many-body initial nuclear environment. In the context of explaining deviations
from binary scaling, the focus is then to understand how a nucleus differs from $\sim200$ nucleons. 
Schematically, differences appear both before and during/after the collision. In the former case
one has to deal with \emph{shadowing}\footnote{Properly speaking, shadowing refers only to the small-$x$ region.}, i.e. the modification of the Parton Distribution Functions 
(PDFs) in the nucleus relative to the the single nucleon, and with \emph{energy loss} of the partons
before the hard collision event. In the latter case a (quasi)formed $Q\overline{Q}$ state may be absorbed
while traversing the nuclear environment.

The treatment of these effects is mostly phenomenological and data driven. Reviews can be 
found in \cite{Brambilla:2010cs,Mocsy:2013syh}. In summary, shadowing is addressed
by parametrizing the nuclear PDFs; quark ones can be constrained from $eA$ deep-inelasting scattering
data, whereas the gluon distribution needs to be inferred indirectly. This introduces a sizeable source
of uncertainty in these parametrizations. Similarly, nuclear absorption introduces uncertainties
in the estimation of the absorpion cross sections for the different states and for their precursors, 
which may also be in a colour octet state. Finally, we also mention that the Color Glass Condensate
(CGC) framework (see \cite{McLerran:2001sr} for a review) can be used to study the initial state effects.

\begin{figure}[ht]
	\begin{center}
 		\includegraphics[width=10cm]{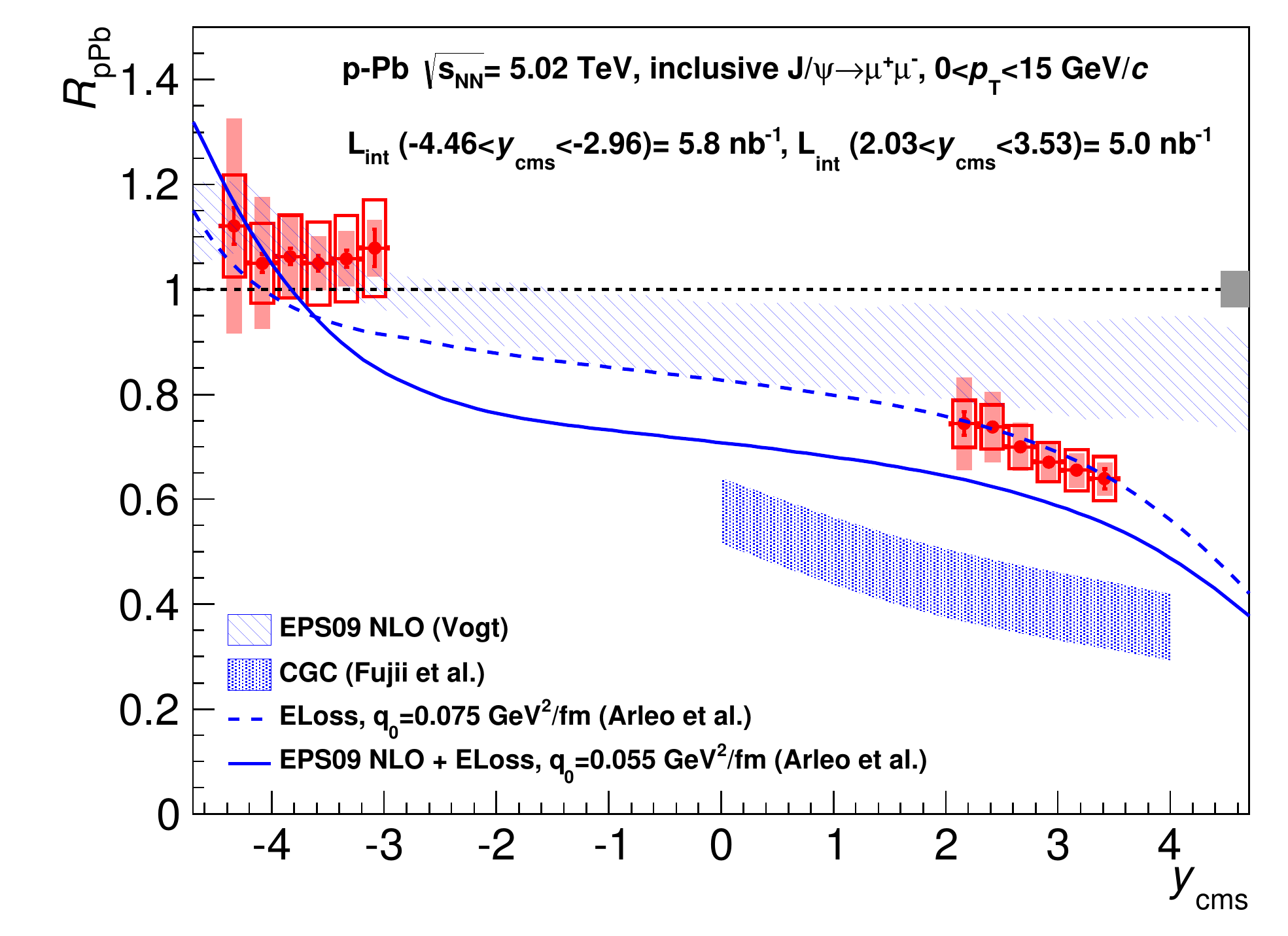}
 	\end{center}
 	\caption{The ALICE data on $J/\psi$ production in $pPb$ collision (relative to $pp$) \cite{Abelev:2013yxa} and its comparison with theoretical calculations 
	of cold nuclear matter effects and the associated uncertainties. We refer to \cite{Abelev:2013yxa} for the references on these theory predictions. Figure taken from \cite{Abelev:2013yxa}. }
 	\label{fig_alice}
 \end{figure}

It is on the other hand necessary to remark that such effects,
 	hence the name, also contribute to $pA$ and $dA$ collisions, where  hot nuclear matter effects 
	are unlikely to occur. Experimental data is available for this
 	asymmetrical collisions at RHIC and LHC and can be used to constrain the theoretical descriptions
 	of these effects. For illustration purposes, Fig.~\ref{fig_alice} shows the recent ALICE
	data on $J/\psi$ production in $pPb$ collisions \cite{Abelev:2013yxa} and its comparison with theoretical calculations 
	of cold nuclear matter effects and the associated uncertainties.
%
\section{Hot nuclear matter effects}
As mentioned in the introduction, we are now dealing with the effects caused by a hot, deconfined
medium, which were those that motivated the initial hypothesis of Matsui and Satz. Their reasoning 
was based on colour screening, a well-known property of plasmas, abelian and non-abelian alike. Since
the screening lenght is approximately inversely proportional to the temperature, in this scenario
one expects a sequential suppression, from the less tightly to the more tightly bound states, as the 
temperature is increased. 

We give here a sketch of some modern, QCD-based field-theoretical approaches to the dynamics
of a $Q\overline{Q}$ pair in the QGP. Due to space limitations, recent developments using other approaches, such as
AdS/CFT, will not be discussed.
\subsection{Extracting the spectral function from the lattice}
\begin{figure}[ht]
	\begin{center}
		\includegraphics[width=8cm]{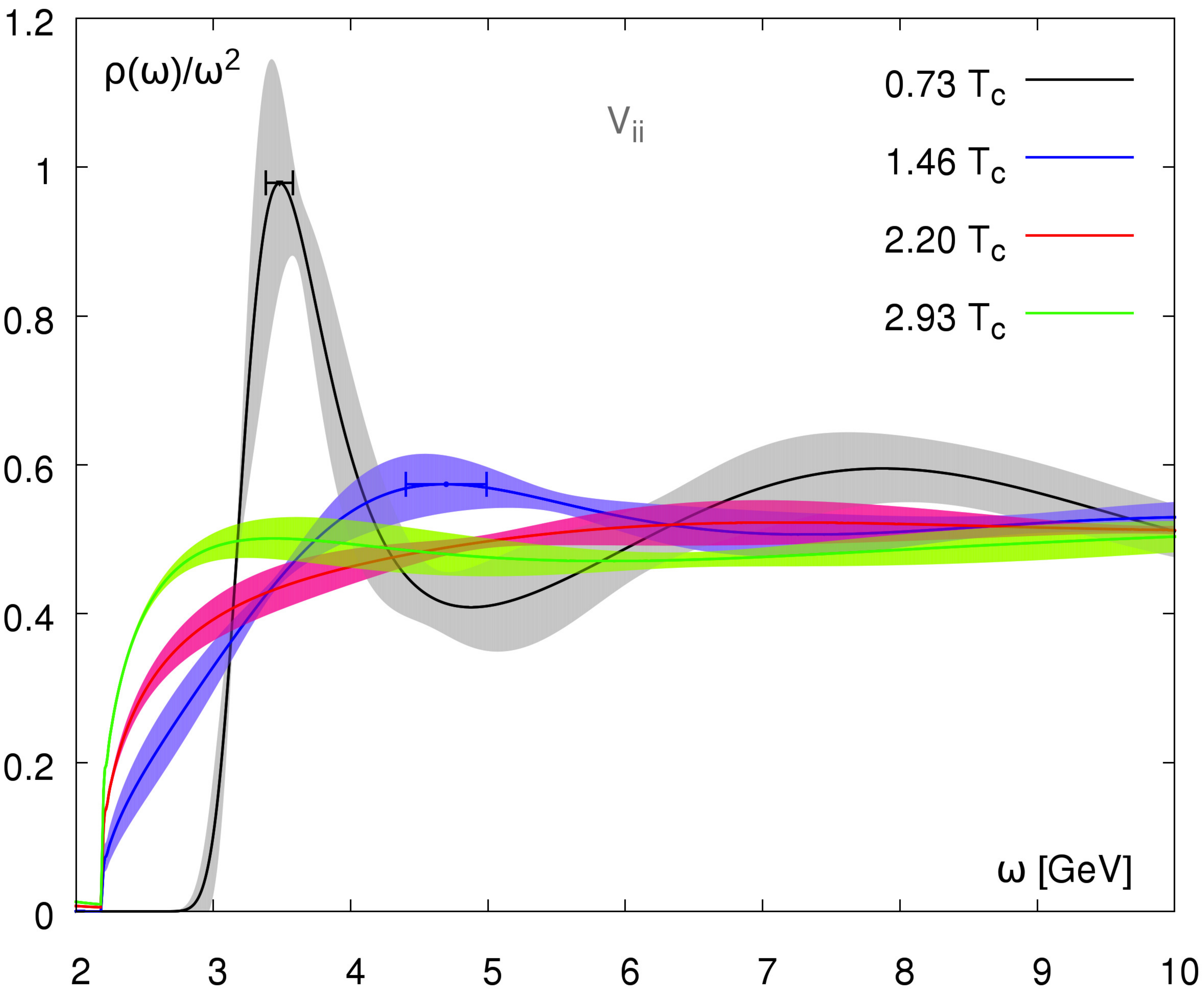}
	\end{center}
	\caption{The MEM charmonium spectral function in the vector channel \cite{Ding:2012sp}. 
	The shaded bands represent the statistical uncertainties only. Figure taken from \cite{Ding:2012sp}.}
	\label{fig_ding}
\end{figure}

All the relevant information about the in-medium bound state is contained in the spectral
function $\sigma(\omega,\bp)$ of the relevant local mesonic operator $J_H(x)\equiv\overline\psi(x)\Gamma_H\psi(x)$,
where $\Gamma_H$ is the appropriate Dirac structure and $\sigma$ is given by the imaginary
part of the Fourier-transformed retarded correlator of $J_H$. Being a Minkowskian quantity,
$\sigma$ is not directly accessible on the lattice. One then exploits the following equation
\begin{equation}
	\label{spf}
	G(\tau,\bp)\equiv\int d^3x e^{i\bp\cdot\bx}\langle J_H(\tau,\bx) J_H(0,{\bf 0})\rangle=
	\int _0^\infty d\omega \sigma(\omega,\bp)\frac{\cosh(\omega(\tau-1/(2T)))}{\sinh(\tau/(2T))},
\end{equation}
which relates the spectral function on the r.h.s. with the Euclidean correlator on the l.h.s. The 
latter can be measured on the lattice, but the extraction of the former requires the inversion of the
above equation through a Bayesian technique known as the Maximum Entropy Method (MEM), as first performed
in \cite{Asakawa:2003re}. Over the years however it was realized how this approach is prone to 
systematic effects introduced by the choice of the priors and the low sensitivity
of the Euclidean correlator in Eq.~\eqref{spf} to changes in the temperature; recent studies
 \cite{Ding:2012sp} find no charmonium 
bound states above 1.5 $T_c$, as shown in Fig.~\ref{fig_ding}. Spatial correlators might also provide a more temperature-sensitive
Euclidean probe \cite{Karsch:2012na}. 
\subsection{Potential models}
\begin{figure}[ht]
	\begin{center}
		\includegraphics[width=8cm]{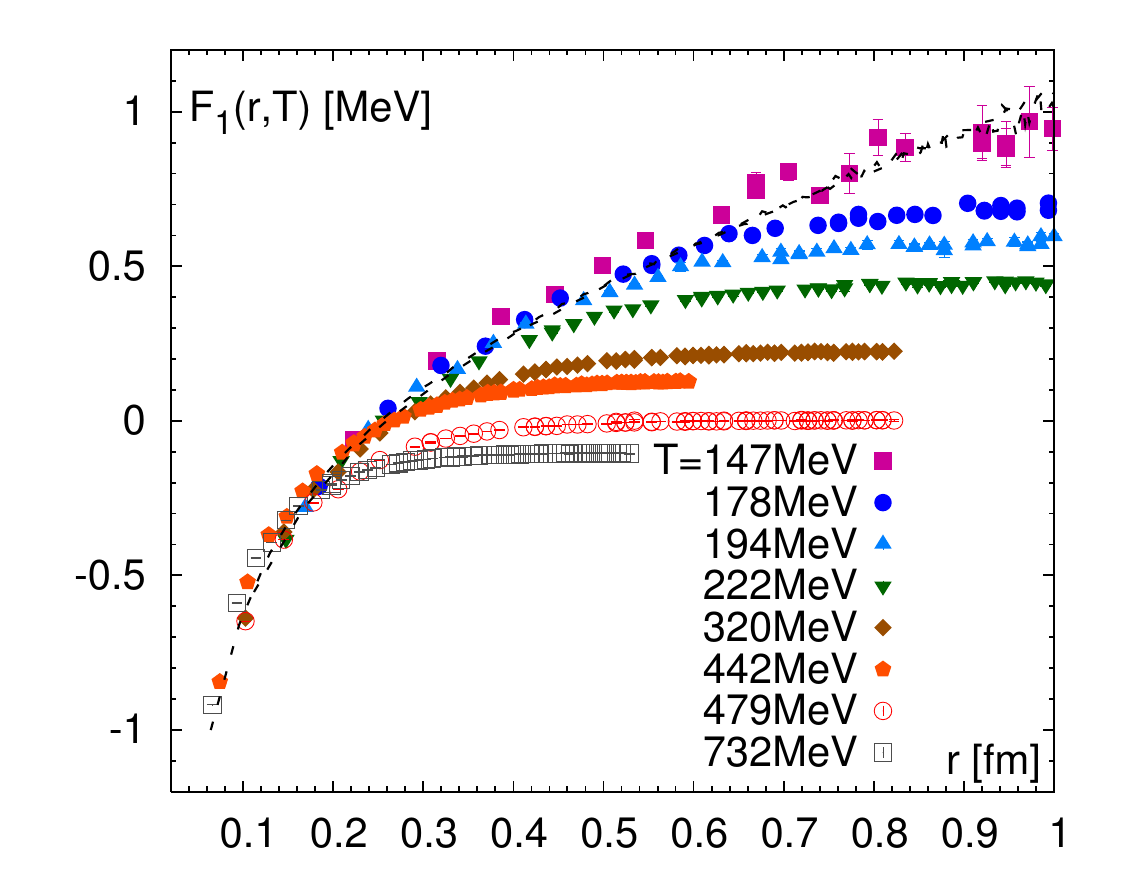}
		\caption{A recent determination \cite{Bazavov:2012fk} of the colour-singlet free energy in Coulomb gauge in $2+1$ flavor QCD. Figure taken from \cite{Bazavov:2012fk}.}
		\label{fig_f1}
	\end{center}
\end{figure}

Potential models have been and are a very popular approach to the problem. First introduced in \cite{Karsch:1987pv}, they rely on the assumption that all in-medium dynamics can be encoded in a 
temperature-dependent potential plugged in a Schr\"odinger equation. A common choice for the potential
is the so-called singlet free energy \cite{Nadkarni:1986as}, a gauge-dependent correlator of two Polyakov lines which can be 
easily measured on the lattice in Coulomb gauge. A recent calculation \cite{Bazavov:2012fk} is plotted
in Fig.~\ref{fig_f1} and indeed shows a pattern of increasing screening with the temperature.
 Other approaches employ the Legendre-transformed
internal energy instead, resulting in a more binding potential, resulting in widely varying phenomenological
implications for the $J/\psi$ and other states. We refer to
 \cite{early_reviews,Mocsy:2013syh} for reviews. Here we just wish to remark that 
these models are not directly derived from QCD, nor is the specific form of the potential to be used. 
 As we shall discuss in the next section, the Effective Field Theory (EFT) approach addresses these issues. 
\subsection{The EFT approach}
This approach is based on the exploitation of scale separations at the Lagrangian level and is inspired
from the successful $T=0$ framework. There, one makes use of the non-relativistic hierachy $m\gg mv\gg mv^2$, 
where $m$ is the heavy quark mass and $v\ll1$ the relative velocity. Upon integrating out the scale $m$
Non-Relativistic QCD (NRQCD) is obtained \cite{nrqcd}.  The second step
 is the integration of the scale $mv$, which leads to potential Non-Relativistic QCD (pNRQCD) \cite{pnrqcd}.
  The relative hierarchical position
of $\lqcd$ and $mv$ establishes whether this integration is to be performed
perturbatively or non-perturbatively. In the former case the $Q\overline{Q}$ sector
of the Lagrangian reads \cite{pnrqcd}
\begin{eqnarray}
{\cal L}_{\textrm{pNRQCD}} &=& 
 \int d^3r \; {\rm Tr} \,  
\Bigl\{ {\rm S}^\dagger \left[ i\partial_0+\frac{\nabla^2}{m}-V_s \right] {\rm S} 
+ {\rm O}^\dagger \left[ iD_0 +\frac{\sd^2}{m}-V_o \right] {\rm O} 
\nonumber \\
&& \hspace{1.8cm}
+  \left( {\rm O}^\dagger \br \cdot g\be \,{\rm S} + \textrm{H.c.} \right)
+ \frac{1}{2} {\rm O}^\dagger \left\{ \br\cdot g\be \,, {\rm O}\right\} 
+ \dots\,\Bigr\} ,
\label{pNRQCD}	
\end{eqnarray}
where $S$ and $O$ are the colour-singlet and colour-octet $Q\overline{Q}$ bilinear field,
which appear as the degrees of freedom of the theory. The second line contains their interactions
at leading (dipole) order in the expansion for $mv^2\ll mv$. At the zeroth order in that expansion
the equation of motion for the singlet field is a Schr\"odinger equation, with the potential
rigorously determined from QCD through the matching procedure. In the strong-coupling case
the octet degrees of freedom are integrated out as well, obtaning a Schr\"odinger picture to all orders.

The extension of this framework to finite temperatures, started in \cite{Brambilla:2008cx,Escobedo:2008sy}
(see \cite{Mocsy:2013syh,Ghiglieri:2013iya} for reviews), requires the introduction of the thermal scales 
(the temperature $T$ and the screening masses) in the problem. In  phenomenologically relevant
situations the heavy quark mass is larger than the temperature, $m\gg T$, hence the first step is unchanged
and yields standard NRQCD. At this point one can either use this theory directly or proceed to obtain
other EFTs. In the former case  NRQCD is simulated on the lattice, which has the advantages of 
making bottomonium physics accessible nonperturbatively and furthermore simplifies the convolution of
the spectral function on the r.h.s. of Eq.~\eqref{spf} to a simpler Laplace transform, which is 
easier to invert with MEM techniques. The results
show a surivival of the $S$ bottomonium ground states up to $\sim 2 T_c$ and a rapid disappearance
of the $P$ states after the transition \cite{nrqcd_lattice}. 

On the other hand, one can proceed to integrate out other scales. If one assumes weak coupling,
then the thermal scales also develop a hierarchy and one has to consider all the relevant scenarios,
from $T\gg mv$ to $T\siml mv^2$ and proceed to systematically integrate out all scales, leaving only $mv^2$
and smaller scales as dynamical.  In all cases, once the scale $mv$ is integrated out, the resulting EFT
resembles pNRQCD and its Lagrangian~\eqref{pNRQCD}, yielding, similarly to the previous $T=0$ discussion,
 a modern and rigorous definition of the potential. A key feature of the potentials obtained
 in the different scenarios is that they are \emph{complex} \cite{Laine:2006ns,Beraudo:2007ky,Brambilla:2008cx}, i.e. they contain a sizeable imaginary part
 that encodes the decoherence effects that interactions with the medium cause; this leads to
  a thermal width for the state. For instance, for distances of the order of the screening length $1/m_D$,
 $m_D^2=g^2 T^2(1+n_f/6)$ being the Debye mass,
  the potential reads \cite{Laine:2006ns}
\begin{equation}
	\label{laine}
	V_s(r\sim 1/m_D) =  -\frac43\,\frac{\als}{r}\,e^{-m_Dr} -\frac43\als m_D
	+ i\frac83\,\als\, T\,\int_0^\infty dt \,\left(\frac{\sin(m_Dr\,t)}{m_Dr\,t}-1\right)\frac{t}{(t^2+1)^2}
	\,.
\end{equation}
The real part is a screened Debye potential (plus a constant, negative self-energy 
 contribution) and is actually smaller, in the power-counting of the theory, than the imaginary part,
 highlighting its importance.
 
  Within the weakly-coupled EFT framework one can then proceed to compute medium
   modifications to the spectra and width (see \cite{Ghiglieri:2013iya} for a review), which have been found by \cite{nrqcd_lattice}
    to be in qualitative agreement with the aforementioned lattice NRQCD results.

\begin{figure}[ht]
	\begin{center}
		\includegraphics[angle=-90,scale=0.165]{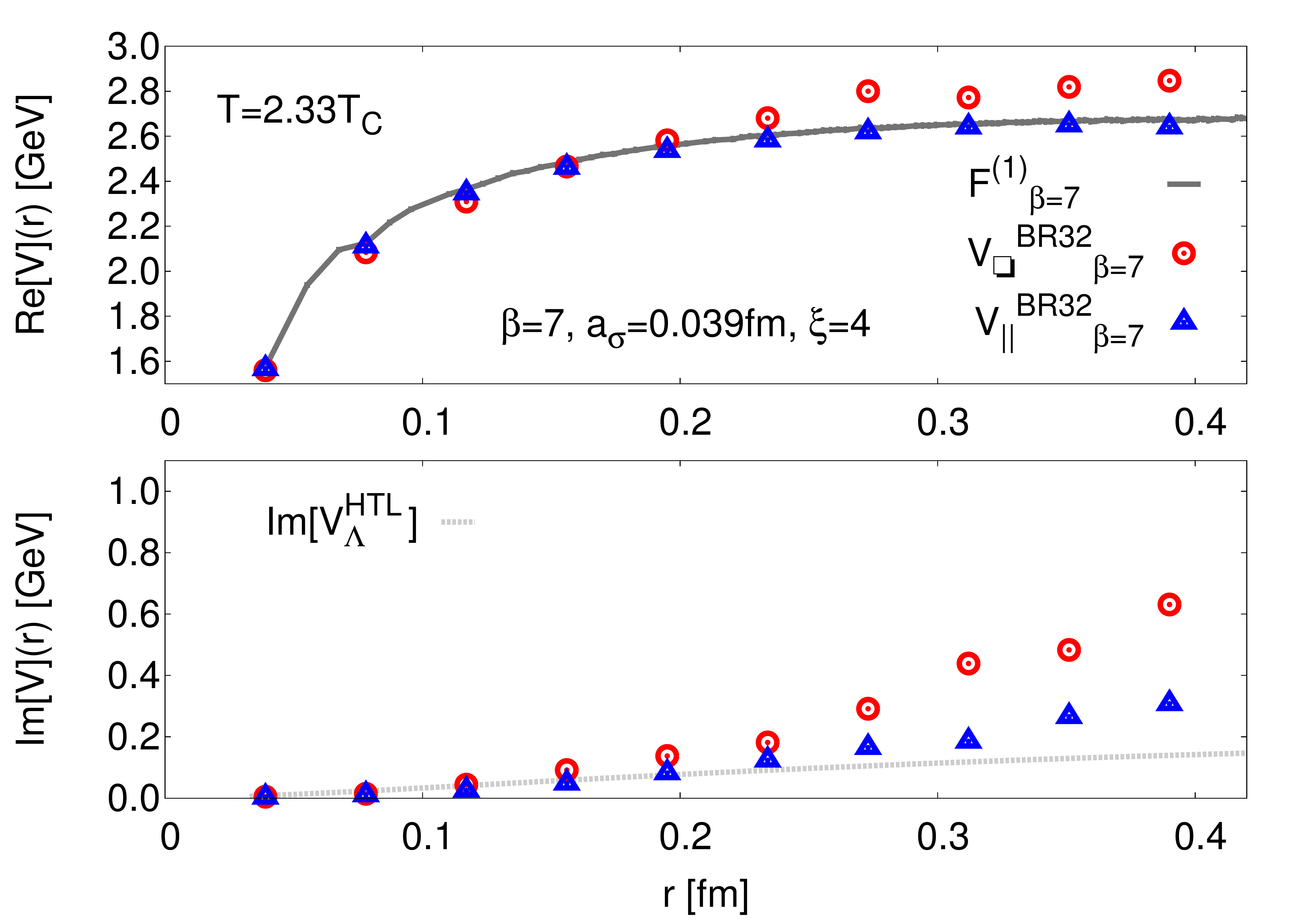}
	\end{center}
	\caption{The real and imaginary parts of the potential obtained in \cite{Burnier:2013nla}. The data
	 points come from different correlators of Wilson lines.
	  $\mathrm{Im}[V^\mathrm{HTL}_\Lambda]$ is
	the perturbative result of \cite{Laine:2006ns}, as given by Eq.~\eqref{laine}. Figure taken from \cite{Burnier:2013nla}.}
	\label{fig_br}
\end{figure}

In the case where no hierarchical separation is present or assumed between $T$, $mv$ and $\lqcd$
one can integrate out all these scales at once, obtaining, as mentioned before, just a Schr\"odinger-like
singlet sector. This requires a non-perturbative determination of the potential. Recent efforts have
shown very promising advancements towards the extraction of the static (infinite mass) complex potential
 from lattice QCD. In Refs.~\cite{rothkopf,Burnier:2013nla} it has been shown how improvements 
 in the MEM can be used
 to obtain a potential whose imaginary part, at short distances, is of the same size of the perturbative
 result \cite{Laine:2006ns}, as shown in Fig.~\ref{fig_br} (a similar approach,
 although not relying on MEM, has been very recenty reported in \cite{peter_hp2013}).
  The results also show a dependence
 on the operator being measured on the lattice: more effort is needed to establish the appropriate ones, 
 also for non-static corrections.

Phenomenologically, the complex potential yields a broadening of the spectral functions at lower $T$ than 
the purely real potential models \cite{lainespf,Petreczky:2010tk} (see \cite{Riek:2010py} for analogous
conclusions in the T-matrix approach). The relation of the imaginary
part with earlier approaches in the literature has been investigated in \cite{Riek:2010py,width}. A 
model implementing an anisotropic complex potential, coupled to the hydrodinamical evolution of the plasma,
has been shown to describe well the LHC $\Upsilon$ data \cite{strickland}, highlighting
the importance of a dynamical description of the medium, while in \cite{CasalderreySolana:2012av}
the possible relevance of vacuum-medium interference effects has been pointed out.
 The impact of the
relative velocity of quarkonium in the plasma has been considered in \cite{moving,nrqcd_moving}. We refer
to \cite{Mocsy:2013syh} for a wider review of the phenomenological applications.
\section{Conclusions}
We have given a brief overview of some theoretical aspects in the description of quarkonia in heavy ion 
collisions. The initial stages of the collision are affected by CNM effects, such as shadowing, initial
energy loss and nuclear absorption. The approach to these issues is mostly phenomenological;
 proton(deuteron)-nucleus collisions represent a very useful tool to this end, as these effects are present
 in those environments too.
 
For what concerns the hot medium effects, we have summarized some QCD-based approaches. The direct
extraction of the $Q\overline{Q}$ spectral function from the lattice is hampered by the need
of an analytical continuation. Nevertheless, qualitative results on dissociation temperatures can be
extracted \cite{Ding:2012sp,Karsch:2012na}. We have then shown how the long-employed potential models can be brought in contact with QCD
using Effective Field Theories \cite{nrqcd,pnrqcd}, which allow for a rigorous QCD derivation of the potential, which turns out
to be complex \cite{Laine:2006ns,Beraudo:2007ky,Brambilla:2008cx}, its imaginary
 part encoding the thermal width caused by interactions with the plasma.
 The potential can be determined in perturbation theory at weak coupling and recent developments are showing
the viability of nonperturbative determinations \cite{rothkopf, Burnier:2013nla}. Furthermore, NRQCD itself can be put on the lattice \cite{nrqcd_lattice},
which is particulary advantageous for bottomonium.

\end{document}